\documentclass[aps,preprint,prl,showpacs,preprintnumbers,amsmath,amssymb]{revtex4-1}

\usepackage{graphicx}
\usepackage{dcolumn}
\usepackage{bm}
\usepackage{txfonts}
\usepackage{color}
\usepackage{pdfpages}

\makeatletter
\AtBeginDocument{\let\LS@rot\@undefined}
\makeatother

\begin{document}
\title{Tuning low-energy scales in YbRh$_2$Si$_2$ by non-isoelectronic substitution and pressure}

\author{M.-H.~Schubert$^{1}$}
\author{Y.~Tokiwa$^{2}$}
\author{S.-H.~H\"{u}bner$^{1}$}
\author{M. Mchalwat$^{1}$}
\author{E. Blumenr\"{o}ther$^{1}$}
\author{H.S.~Jeevan$^{1}$}
\author{P. Gegenwart$^{2}$}
\affiliation{$^{1}$I. Physik. Institut, Georg-August-Universit\"{a}t G\"{o}ttingen, D-37077
G\"{o}ttingen\\$^{2}${Experimental Physics VI, Center for Electronic Correlations and Magnetism, University of Augsburg, 86159 Augsburg, Germany}}

\date{\today}


\begin{abstract}
The heavy-fermion metal YbRh$_2$Si$_2$ realizes a field-induced quantum critical point with multiple vanishing energy scales $T_{\rm N}(B)$ and $T^\ast(B)$. We investigate their change with partial non-isoelectronic substitutions, chemical and hydrostatic pressure. Low-temperature electrical resistivity, specific heat and magnetic susceptibility of Yb(Rh$_{1-x}$T$_x$)$_2$Si$_2$ with T=Fe or Ni for $x\leq 0.1$, magnetic fields $B\leq 0.3$~T (applied perpendicular to the c-axis) and hydrostatic pressure $p\leq 1.5$~GPa are reported. The data allow to disentangle the combined influences of hydrostatic and chemical pressure, as well as non-isoelectronic substitution. In contrast to Ni- and Co-substitution, which enhance magnetic order, Fe-substitution acts oppositely. For $x=0.1$ it also completely suppresses the $T^\ast$ crossover and eliminates ferromagnetic fluctuations. The pressure, magnetic field and temperature dependences of $T^\ast$ are incompatible with its interpretation as Kondo breakdown signature.
\end{abstract}
\maketitle

Quantum phase transitions are of central importance in correlated materials. Whether well-defined quasiparticles exist at a quantum critical point (QCP) is relevant for understanding cuprates and heavy-fermion metals~\cite{Ramshaw,Custers}. The latter consist of lattices of certain f-electrons and realize quantum criticality arising from two competing interactions: the indirect exchange coupling (RKKY interaction) between the f-electrons, mediated by conduction electrons and the Kondo screening acting on each f-moment site. Since both interactions depend on the antiferromagnetic (AF) exchange $J$ between f- and conduction electrons, quantum criticality is realized by tuning $J$ with pressure, chemical substitution or magnetic field~\cite{Lohneysen,Gegenwart08}. Several heavy-fermion metals have been studied near QCPs, revealing non Fermi liquid (NFL) behavior, as well as the occurrence of unconventional superconductivity~\cite{Mathur}.

Tetragonal YbRh$_2$Si$_2$ with very weak AF ordering at $T_{\rm N}=70$~mK is one of the best studied model systems for quantum criticality in heavy-fermion metals~\cite{Trovarelli}. By application of a small critical magnetic field $B_{\rm c}=0.06$~T ($B\perp c$) it displays a QCP and paramagnetic Fermi liquid behavior occurs at $B>B_{\rm c}$~\cite{Gegenwart02}. Hydrostatic pressure~\cite{Tokiwa09} or chemical pressure~\cite{Friedemann}, induced by few atomic \% substitution of Rh by isoelectronic Co, stabilizes the AF ordering and enhances the critical field $B_{\rm c}$. Vice versa, volume expansion induced by negative chemical pressure weakens the AF ordering and reduces $B_{\rm c}$~\cite{Custers,Friedemann}. Close to the field-induced QCP a quasi-linear temperature dependence of the electrical resistivity and divergences of the specific heat coefficient~\cite{Gegenwart02}, magnetic susceptibility~\cite{Gegenwart05} and Gr\"uneisen parameters~\cite{Kuechler,Tokiwa} indicate NFL properties that are incompatible with the theory of itinerant AF quantum criticality~\cite{Millis}.

Measurements of the isothermal field dependence of the Hall coefficient, magnetoresistance, magnetostriction, magnetization and entropy have revealed a crossover scale labeled $T^\ast(B)$, or $B^\ast(T)$~\cite{Paschen,Gegenwart07,Friedemann10}. This crossover is independent from the boundary of the AF order $T_{\rm N}(B)$ and the Fermi-liquid crossover in electrical resistivity, although it merges for undoped YbRh$_2$Si$_2$ at ambient pressure at lowest measured temperatures the critical field $B_{\rm c}$. Since $T^\ast(B)$ increases with increasing field and the magnetic susceptibility $\chi(T)$ for $B\perp c$ displays a local maximum at $T^\ast$, the crossover has been associated with a partial ferromagnetic (FM) polarization~\cite{Gegenwart05}. Subsequently, the crossover has been interpreted as signature of a Fermi surface reconstruction due to a Kondo breakdown for the following reasons: (i) it appears in the Hall coefficient but for $B\parallel c$ it cannot be related to an anomalous Hall effect~\cite{Paschen} and (ii) its widths in various physical properties displays a linear temperature dependence~\cite{Friedemann10}. Extrapolation to $T=0$ therefore suggests jumps of the Hall coefficient and magnetoresistance, which were taken as evidence for a transition from small to large Fermi surface at the QCP~\cite{Paschen,Gegenwart07,Friedemann10}, in agreement with the expectation for the Kondo breakdown scenario~\cite{Si}. Alternatively, $T^\ast$ was related to spin-flip scattering of critical quasiparticles~\cite{WA} or a Zeeman-driven narrow-band Lifshitz transition~\cite{Hackl}. The latter scenario requires, however, fine-tuning of a very narrow peak in the density of states~\cite{Hackl_comment}.

Remarkably, $T^\ast(B)$ is only very weakly influenced by hydrostatic or (negative) chemical pressure, induced by partial isovalent substitutions, which enlarges or diminishes the magnetically ordered phase, leading to either $B^\ast<B_{\rm c}$ or $B^\ast>B_{\rm c}$, respectively~\cite{Tokiwa09,Friedemann}. This was ascribed to an itinerant QCP at $B_{\rm c}$ in the former and a spin-liquid regime in between $B_{\rm c}$ and $B^\ast$ in the latter case~\cite{Friedemann}. However, this interpretation relies on the assumption that $B^\ast$ indeed indicates a change from small to large Fermi surface volume, due to a Kondo breakdown. In fact it is surprising, that a Kondo breakdown would be so weakly influenced by pressure or (negative) chemical pressure, which tunes the balance of the Kondo to the RKKY interaction. More recently the Kondo breakdown interpretation of $T^\ast$ was also questioned by high-resolution ARPES. At zero-field and temperatures down to 1~K it detected a large Fermi surface~\cite{Kummer}. However, it was argued, that this temperature is still too large to observe a small Fermi surface expected within the Kondo breakdown scenario for $B<B^\ast$~\cite{Paschen16} and a similar rationale was used for the absence of a significant change of scanning tunneling spectroscopy at $B^\ast$, even at 0.3~K~\cite{Seiro}. Thus, further experimental work on the nature of $T^\ast$ at $T<0.3$~K is badly needed.

Below, we report drastic changes of $T^\ast$ by partial non-isoelectronic Fe- or Ni-substitutions, which cannot be related to the effect of chemical pressure. A complete suppression of the crossover scale is found for 10\% Fe-doping and this suppression correlates with the disappearence of low-temperature FM fluctuations. Furthermore, we find for all studied single crystals, that the $B^\ast$ crossover widths depart from a linear $T$ dependence and do not extrapolate to zero for $T\rightarrow 0$. Altogether, the results question that $B^\ast$ for $B\perp c$ is related to a Kondo breakdown and suggest that it results from a partial polarization of fluctuating moments~\cite{Gegenwart05}.

Various flux grown Yb(Rh$_{1-x}$T$_x$)$_2$Si$_2$ single crystals with T=Fe and Ni for $x\leq 0.1$ were characterized and investigated, see supplemental material (SM)~\cite{SM,Mchalwat,Blumenroether,Schubert,Krellner_PhD,Krellner_PhilMag,Klinger,Rossi,Noakes,Bara,Mori,Mederle,Gegenwart_NJP,Hubner,Friedemann_PhD}. The actual doping concentrations, which in some cases deviate from the starting compositions, have been determined with $\Delta x\sim 0.01$ precision. For low-noise alternating current electrical resistivity and magnetic susceptibility measurements down to 20~mK commercial low-temperature transformers were utilized in a dilution refrigerator. Measurements at enhanced temperatures were conducted in the PPMS and MPMS. All experiments were carried out for $B\perp c$. For hydrostatic pressure experiments up to 1.5 GPa a piston cylinder pressure cell with daphne oil as pressure medium has been utilized. The pressure has been determined with a superconducting lead manometer. 

In the following, we discuss the influence of different non-isoelectronic substitutions on the low temperature properties of Yb(Rh$_{1-x}$T$_x$)$_2$Si$_2$ in comparison with previous results of isoelectronic substitution (T=Co and Ir)~\cite{Friedemann} as well as hydrostatic pressure~\cite{Tokiwa09}. Considering the periodic table, Ni- or Fe-doping enhances or reduces the number of conduction electrons in YbRh$_2$Si$_2$, respectively. Additionally, these substitutions also induce chemical pressure and enhance the residual resistivity $\rho_0$, e.g. to 1.8 and 7.6~$\mu\Omega$cm for 3 and $7.5\%$ Fe doping, respectively, similar as for respective Co-substitution~(SM). For undoped YbRh$_2$Si$_2$ an enhancement of $\rho_0$ from $0.5\,\mu\Omega\text{cm}$ to $3~\mu\Omega$cm does not change $T_{\rm N}$, $B_{\rm c}$ or $T^\ast$~\cite{Trovarelli,Krellner_Violation-of_critical_universality}. Furthermore, the hydrostatic pressure dependence of these properties for undoped YbRh$_2$Si$_2$ perfectly matches with the effect of chemical pressure in Yb(Rh$_{1-x}$Co$_x$)$_2$Si$_2$ single crystals, for which latter the residual resistivity is enhanced, e.g. to $10.7\mu\Omega\text{cm}$ for $x=0.07$~\cite{Tokiwa09,Klinger}. Thus, for Fe-doping the observed changes of the low-energy scales reported below are not primarily caused by the effect of disorder but rather result from chemical pressure and non-isoelectronic substitution.

\begin{figure}[t]
\includegraphics[width=0.9\linewidth,keepaspectratio]{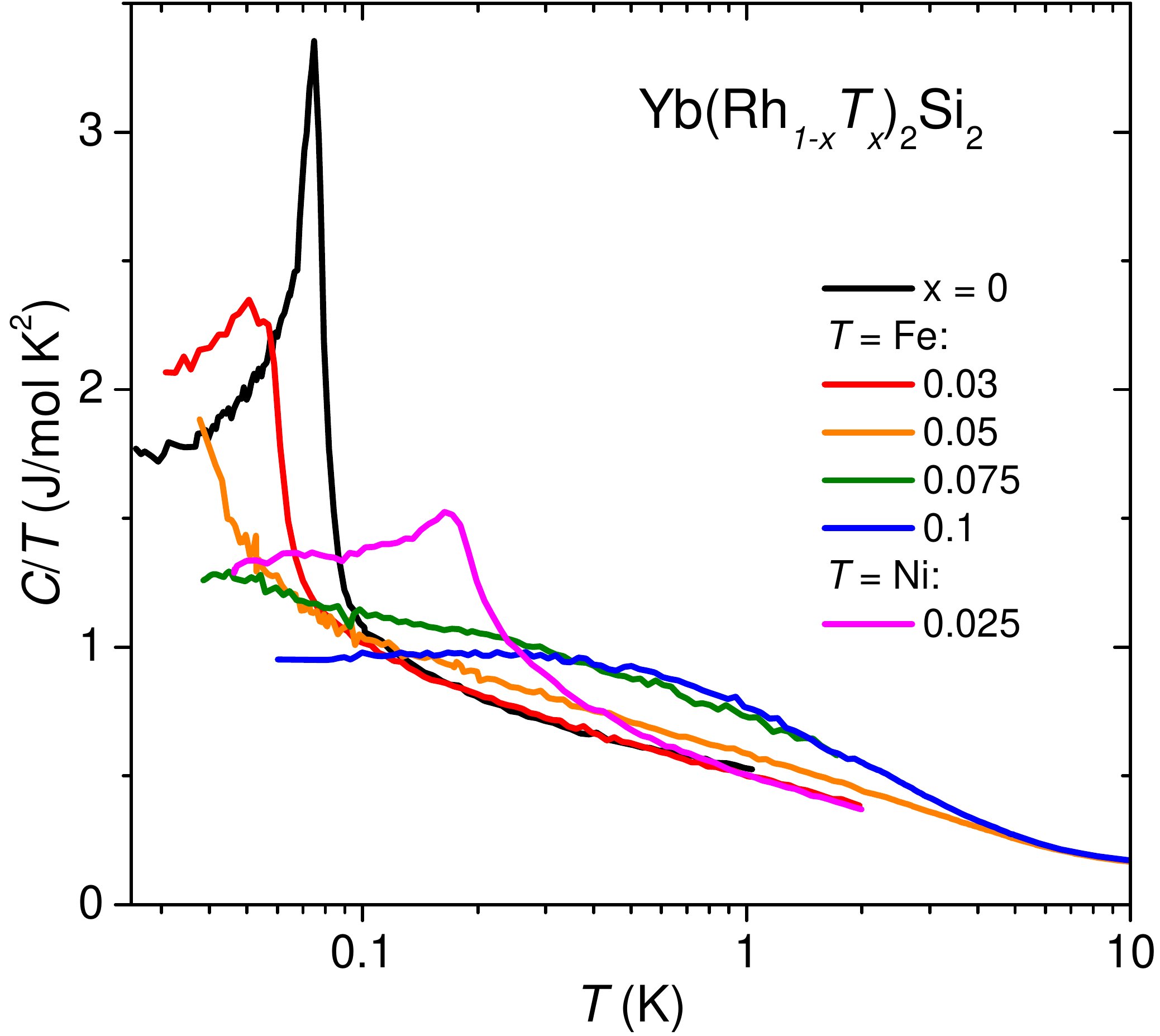}
\caption{Temperature dependence of the specific heat coefficient $C/T$ (on a logarithmic scale) for various single crystals of Yb(Rh$_{1-x}$T$_x$)$_2$Si$_2$ (T=Fe, Ni) with compositions as indicated by the labels. Data for $x=0$ were taken from~\cite{Gegenwart02}.}
\label{fig:Specific_Heat}
\end{figure}

Fig.~1, displaying the low-temperature specific heat coefficient of various investigated single crystals, provides an overview on the tunability of the ground state by doping. The sharp peak at 75~mK for the undoped material indicates the AF phase transition~\cite{Gegenwart02}. Already small substitution with Fe or Ni significantly shifts the transition to lower and higher temperatures, respectively. This cannot be related to the effect of chemical pressure, which acts similarly for partial Fe, Ni and Co substitution, as evidenced by a similar evolution of the lattice constants, resistivity maximum temperature and Kondo temperature (see SM). We therefore associate the disparate change of $T_{\rm N}$ with partial Fe and Ni substitution to non-isoelectronic substitution. For $5\%$ Fe substitution, the steep upturn below 80~mK indicates an AF transition temperature around 40~mK, while $C/T$ for $7.5\%$ Fe follows a logarithmic divergence to lowest temperatures. For $10\%$ Fe substitution a clear saturation of $C/T$ below 0.3~K highlights a paramagnetic Fermi liquid (FL) ground state. This evolution is unexpected, given that (i) similar chemical pressure for Co-substitution enhances $T_{\rm N}$ and (ii) the isostructural end-member YbFe$_2$Si$_2$ orders magnetically at 0.75~K~\cite{Hodges}. As detailed in SM, a non-monotonic evolution of the lattice constants and change of the CEF ground state wave function is expected at $x$ (Fe) $>0.1$.

\begin{figure}[t]
\includegraphics[width=0.9\linewidth,keepaspectratio]{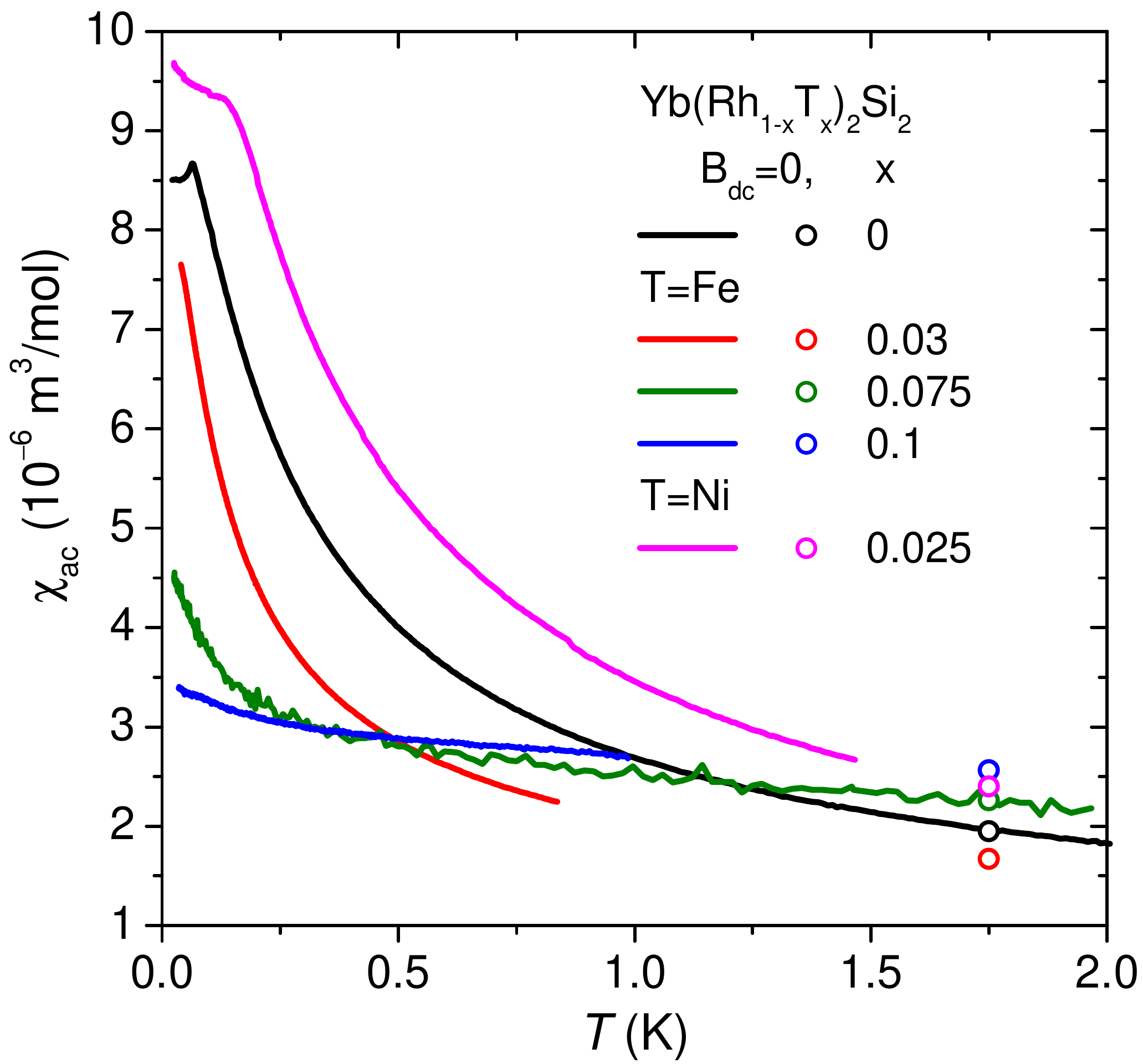}
\caption{Temperature dependence of the magnetic ac-susceptibility (in zero dc-field) for various single crystals of Yb(Rh$_{1-x}$T$_x$)$_2$Si$_2$ with compositions as indicated by the labels. All data have been taken for fields applied within the easy plane perpendicular to the c-axis. The open circles indicate $M/B$ data determined in a SQUID magnetometer (see SM). Data for $x=0$ were taken from~\cite{Gegenwart02}.}
\label{fig:Susceptibility}
\end{figure}

The influence of Fe- or Ni-substitutions on the low-temperature magnetic susceptibility is shown in Fig.~2. The AF phase transition for undoped and Ni-doped crystals results in sharp anomalies. The observed increase of the susceptibility upon cooling, following a $\chi(T)\sim T^{-0.6}$ divergence between 0.3 and 10~K (for undoped and $5\%$ Ge-doped YbRh$_2$Si$_2$) has previously been ascribed to FM fluctuations that compete with AF fluctuations close to $T_{\rm N}$~\cite{Gegenwart05,Gegenwart_JPSJ}. Fe-doping reduces the low-temperature susceptibility by a factor three for $x=0.1$ compared to undoped, 6\% Ir- and 7\% Co-substituted YbRh$_2$Si$_2$~\cite{Gegenwart05,Friedemann}. This indicates a drastic suppression of the Sommerfeld Wilson ratio and thus the FM fluctuations by Fe-doping.

As detailed in SM, we fitted the magnetic susceptibility between 2 and 4~K by a Curie-Weiss law. For undoped YbRh$_2$Si$_2$ it amounts to $-3$~K and increases towards zero for Ni doping in a similar way as previously found for Co-doping~\cite{Klinger} (note, that Co-substitution tunes the ground state towards ferromagnetism~\cite{Lausberg,Hamann}). By contrast, Fe-substitution strongly enhances the negative $\Theta_{\rm W}$ in accordance with a suppression of the FM fluctuations.

Next, we investigate the evolution of the $T^\ast(B)$ crossover with doping. To determine $T^\ast(B)$, we utilize $\chi(T)$ (see SM) and the isothermal magnetoresistance. Similar as previously done for undoped YbRh$_2$Si$_2$, the characteristic crossover field $B^\ast$ is obtained by fitting of $\rho(B)$ at different temperatures to an empirical crossover function, which also allows to determine the temperature dependence of the full width at half maximum (FWHM) of the crossover~\cite{Paschen,Gegenwart07,Friedemann10}. As shown in Fig.~3(a) for selected doped single crystals, the crossover field marks an inflection point in the negative magnetoresistance, in accordance with previous results for pure YbRh$_2$Si$_2$~\cite{Gegenwart07}. While Ni-doping enhances $B^\ast$, the latter is reduced by Fe-doping (cf. the shift of the arrows with increasing Fe substitution in Fig.~3(a)). In addition, also the size of the negative magnetoresistance contribution is drastically reduced. For $10\%$ Fe substitution, the anomaly associated with $B^\ast$ has completely disappeared and a featureless positive magnetoresistance behavior is found.

\begin{figure}[t]
\includegraphics*[width=0.5\linewidth,keepaspectratio]{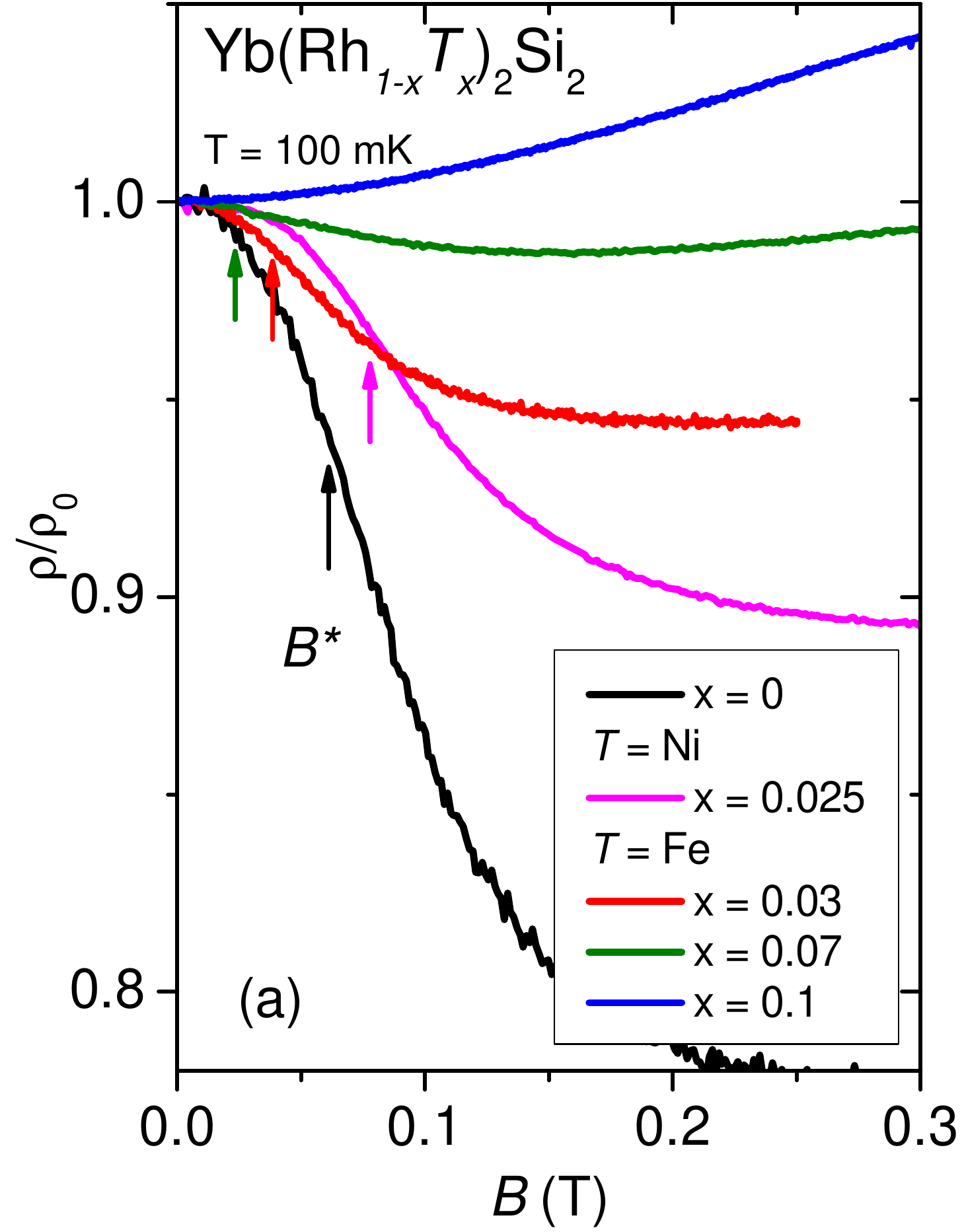}
\includegraphics*[width=0.47\linewidth,keepaspectratio]{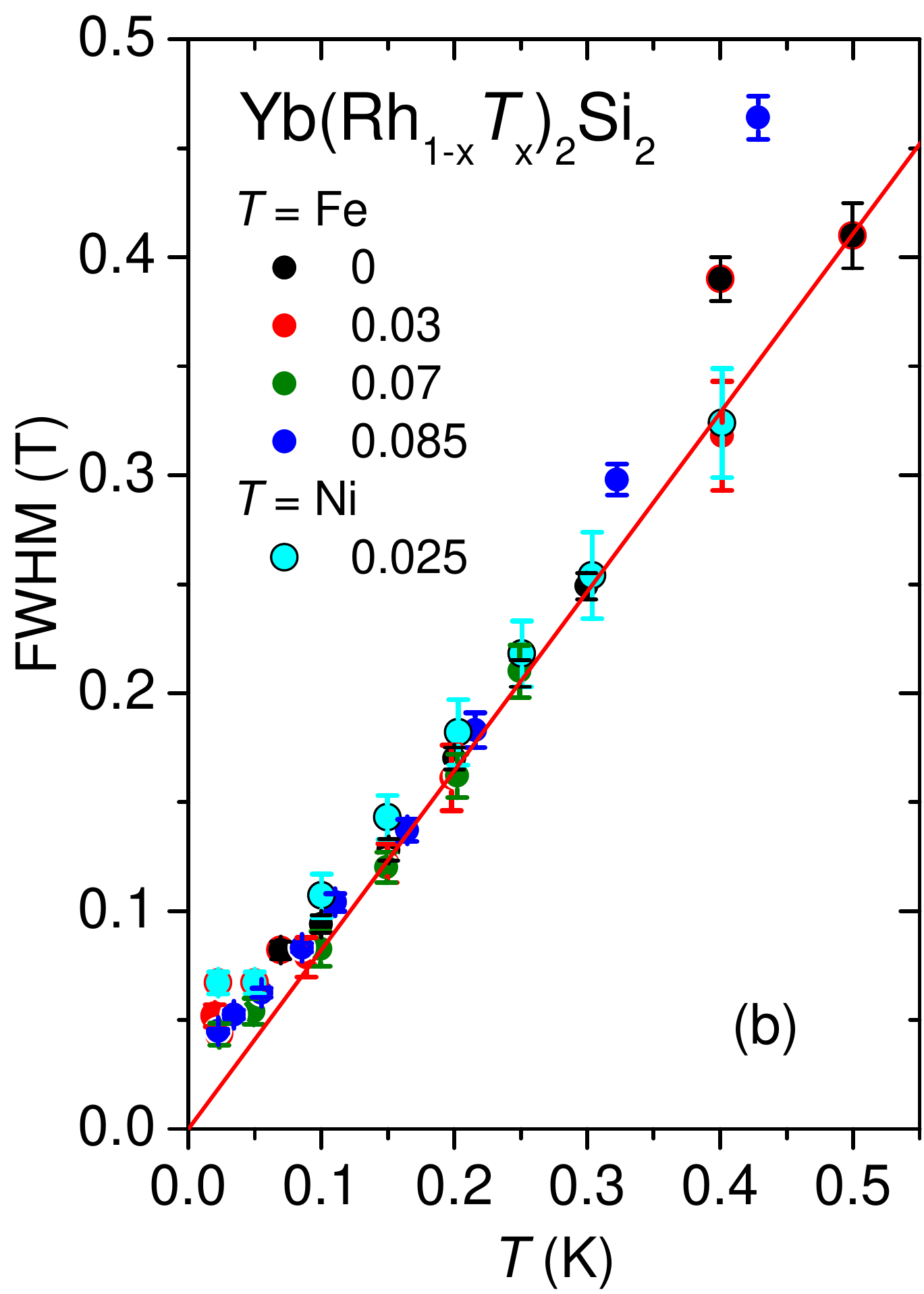}
\caption{Normalized isothermal magnetoresistance of various Yb(Rh$_{1-x}$T$_x$)$_2$Si$_2$ single crystals at 100~mK (a). Arrows mark $B^\ast$ determined by fitting the data to an empirical crossover function of the form $f(B,T)=A_2-(A_2-A_1)/[1+(B/B^\ast(T))^p]$~\cite{Paschen,Gegenwart07,Friedemann10}. Temperature dependence of the respective FWHM values (b). The red line indicates a linear temperature dependence and is drawn identically to ~\cite{Friedemann10}.}
\label{fig:Restivity}
\end{figure}

Fig.~3(b) displays the temperature dependence of the FWHM for the various doped single crystals as determined from magnetoresistance measurements shown in panel (a). At temperatures above 0.1~K, the data agree very well with the same linear dependence (red line) found in previous magnetoresistance and Hall effect measurements up to 1~K~\cite{Friedemann10}. Note, that the FWHM at 0.1~K is still large, i.e., around 0.08~T, indicating that even at this low temperature, this is a rather broad crossover. Clearly for all our studied systems the FWHM is above the red line below 0.1~K. Extrapolation to a zero crossover width at $T=0$ is thus invalid. A similar trend is visible in most data sets from~\cite{Friedemann10} for $x=0$ (see SM) and for Yb(Rh$_{0.93}$Co$_{0.07}$)$_2$Si$_2$~\cite{Brando13}, where the deviation is found at temperatures below $T_N$. Most importantly, we observe deviation from a linear $T$-dependence also for systems in which $B^\ast$ does not cross a magnetic phase boundary. One may argue that disorder introduced by chemical substitution leads to an extrinsic additional temperature independent offset to the FWHM. However, we do not even see an increase of the FWHM with tenfold increasing residual resistivity for the systems shown in Fig. 3(b).

\begin{figure}[t]
\includegraphics[width=0.9\linewidth,keepaspectratio]{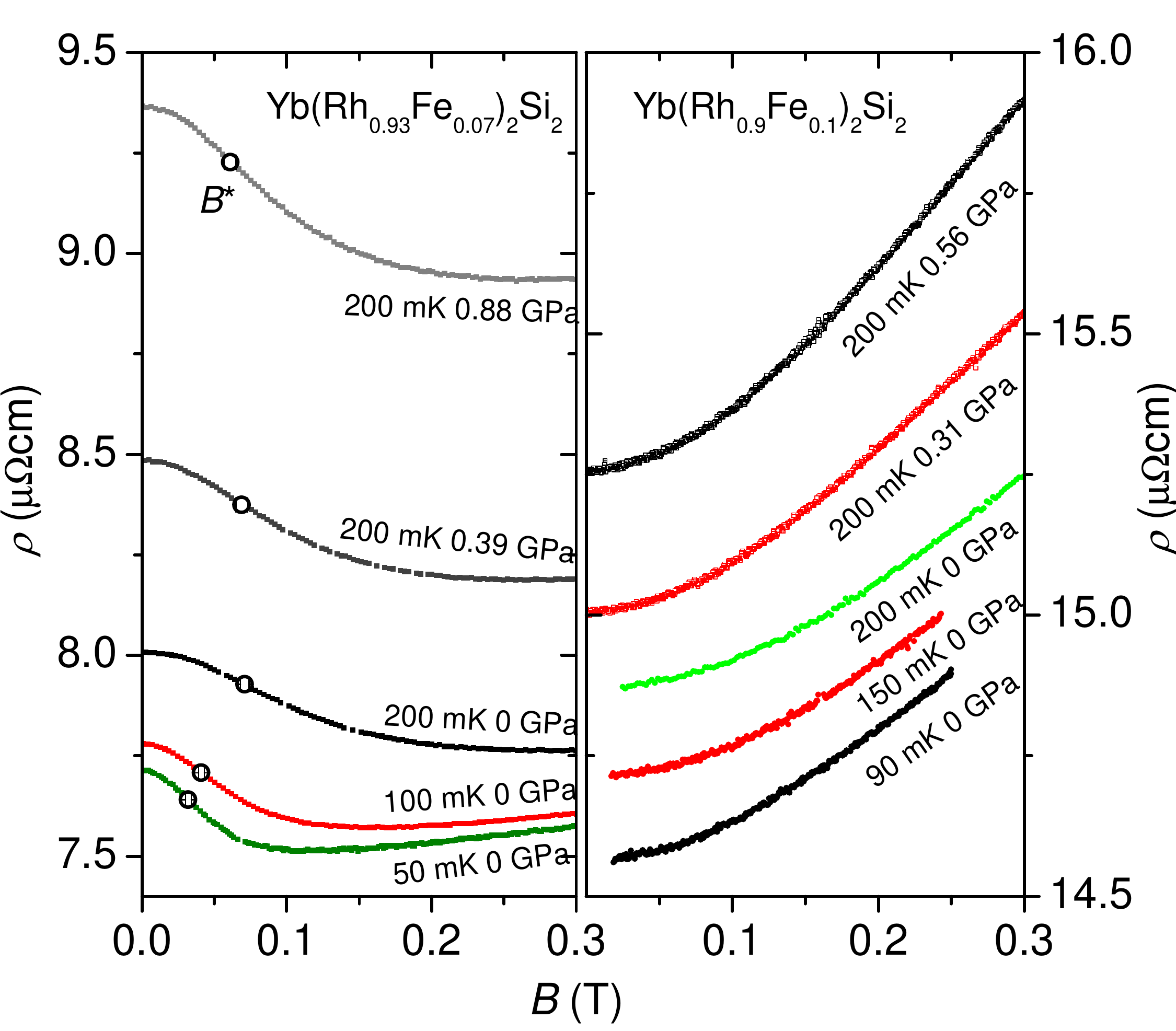}
\caption{Isothermal magnetoresistance of Yb(Rh$_{1-x}$Fe$_x$)$_2$Si$_2$ for $x=0.07$ (a) and $x=0.1$ (b) at various temperatures and pressures. The circles in (a) indicate the position of $B^\ast(T)$.}
\label{fig:Restivity}
\end{figure}

To disentangle the combined effect of chemical pressure and charge carrier doping on $T^\ast(B)$ in Yb(Rh$_{1-x}$Fe$_x$)$_2$Si$_2$, we also performed hydrostatic pressure experiments on $x=0.07$ and $x=0.1$ single crystals. We selected these two concentrations, because the former still shows a tiny crossover signature, while it is fully suppressed for the latter. As discussed above, for undoped YbRh$_2$Si$_2$ as well as partial isoelectronic Co and Ir substitution, the $T^\ast(B)$ scale is almost insensitive of pressure~\cite{Friedemann,Tokiwa09}. As shown in Fig.~4(a), the same holds true for the $7\%$ Fe-doped case. The crossover field $B^\ast$ at 0.2~K shifts only very weakly towards smaller values for pressures up to 0.88~GPa. On the other hand, hydrostatic pressure of this size leads to a rapid stabilization of magnetic order. This is evident from a clear dip in the temperature derivative of the electrical resistivity $d\rho/dT$ at $T_{\rm N}$, shown in the inset of Fig.~5, which is first visible at 0.39~GPa. $T_{\rm N}$ increases with increasing pressure. Isothermal magnetoresistance measurements under hydrostatic pressure are shown in SM. The phase diagram in Fig.~5 summarizes the anomalies from temperature (circles) and field (diamonds) sweeps. While at ambient pressure no magnetically ordered state can be detected, for a pressure of 0.39~GPa the observed Neel temperature and critical field are close to that for undoped YbRh$_2$Si$_2$ at $p=0$. With increasing pressure, the AF phase boundary is enlarged while the $T^\ast(B)$ line does not change much. This is qualitatively similar as found previously for Yb(Rh$_{1-x}$Co$_x$)$_2$Si$_2$ ~\cite{Friedemann} and pressurized YbRh$_2$Si$_2$~\cite{Tokiwa09}. Thus, pressure is able to tune back the AF state, which has been depressed below 40~mK but only weakly changes $T^\ast(B)$.

\begin{figure}[t]
\includegraphics[width=0.9\linewidth,keepaspectratio]{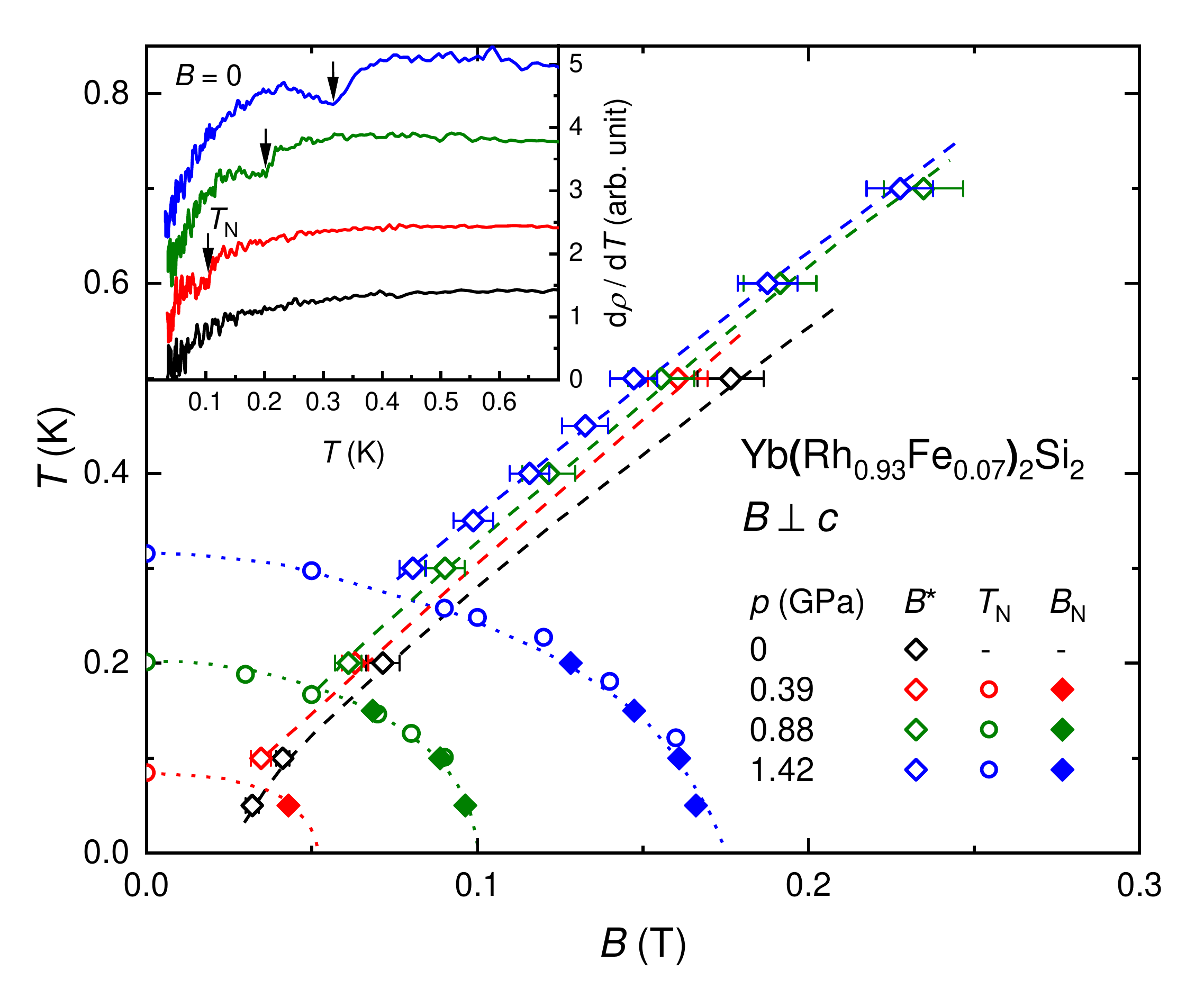}
\caption{Temperature field phase diagram for Yb(Rh$_{0.93}$Fe$_{0.07}$)$_2$Si$_2$ at various hydrostatic pressures. Circles and diamonds determined from temperature and field sweeps, respectively, denote AF phase boundaries and $T^\ast(B)$ crossover as indicated by labels. The Inset shows the signature of $T_{\rm N}$ in the temperature derivative of the electrical resistivity for different pressures. Curves are shifted upwards for clarification. Arrows indicate $T_{\rm N}$.}
\label{fig:Phase-diag-under-pressure}
\end{figure}

Next we discuss the pressure experiments on Yb(Rh$_{0.9}$Fe$_{0.1}$)$_2$Si$_2$. At ambient pressure, this sample is located on the paramagnetic side of the QCP (cf. Fig. 1), shows $T^2$ behavior in zero-field $\rho(T)$ (see SM) and also the $T^\ast(B)$ crossover is completely suppressed and instead a positive magnetoresistance is found (Fig. 3). Since $x=0.1$ is so close to $x=0.085$ for which $T^\ast$ was still observed, we were interested, whether pressure can recover the crossover for $x=0.1$. However, even up to the maximal pressure of 1.37 GPa (see SM and Fig.~4(b)), magnetoresistance remains positive and the $T^\ast$ crossover remains absent. Such pressure has strong influence on undoped YbRh$_2$Si$_2$ or on the $x=0.07$ sample. Therefore, the balance between Kondo and RKKY interaction must significantly be modified by 1.37 GPa. Since this system is so close to the concentration at which the $T^\ast$ crossover has been suppressed, it is therefore fully unexpected within the Kondo breakdown scenario, that such significant pressure is unable to recover the change of the Fermi surface volume. Thus, these data seem incompatible with the notion that $T^\ast$ indicates a Kondo breakdown. Rather than depending on pressure or chemical pressure, $T^\ast$ appears to be highly sensitive to Fe- or Ni-doping.

Another import observation is, that the size of the magnetoresistance crossover disappears before the crossover field $B^\ast$ approaches zero (see SM). Thus, this anomaly is a field-induced effect. The zero-field temperature-pressure/doping plane has no $T^\ast$ signature, in contrast to the general expectation for local quantum criticality, where a Kondo breakdown should also occur in the absence of a magnetic field~\cite{Si}.

While the FM fluctuations are rather robust under positive or negative chemical pressure, Fe-doping leads to a suppression of FM fluctuations (cf. Fig. 2), which coincides with the complete disappearance of the $T^\ast$ crossover. This crossover thus marks a field-induced partial polarization of moments~\cite{Gegenwart05}, indicated by an inflection point of the entropy $S(B)$ at $T^\ast$~\cite{Tokiwa}, which naturally explains the negative magnetoresistance. In fact such signature of moment polarization in susceptibility and magnetoresistance is expected in any metallic magnet by the Zeeman effect. Recently it has e.g. been reported in NbFe$_2$~\cite{Rauch}, YbNi$_4$P$_2$~\cite{Lausberg}, Ce$_3$Pd$_{20}$Si$_6$~\cite{Custers12} and CePdAl~\cite{Zhang}, although in the two latter cases it was interpreted as finite temperature signature of a Kondo breakdown.

Despite the chemical pressure, induced by Fe-doping and the fact that pure YbFe$_2$Si$_2$ is an AF with $T_{\rm N}=0.75$~K~\cite{Hodges} the partial substitution of Rh with Fe suppresses AF order in Yb(Rh$_{1-x}$Fe$_x$)$_2$Si$_2$. For x=0.1 a stable Fermi liquid state develops, which lacks any $T^\ast$ crossover. Isoelectronic substitution of Rh by Co or Ir, corresponds to positive or negative chemical pressure and does not modify the $T^\ast(B)$ crossover line. On the other hand, non-isovalent Fe- and Ni-substitutions depress and enlarge $T^\ast(B)$, respectively.

The previous interpretation of $T^\ast(B)$ (for $B\perp c$ and $B\parallel c$) in YbRh$_2$Si$_2$ as finite-temperature signature of a Kondo breakdown is questioned by the following observations when tuning $T^\ast(B)$ by non-isoelectronic substitutions: 1) the crossover width does not extrapolate to zero,  2) the $T^\ast$ anomaly requires a finite magnetic field, 3) in contrast to the Kondo temperature, it is almost insensitive to pressure and 4) it coheres with FM fluctuations, i.e., once the latter are depressed, $T^\ast$ disappears.

We thank M.~Brando, J.~Dong, S.~Friedemann, C. Geibel, S.~Lausberg, Q.~Si, C. Stingl, M. Vojta and K. Winzer for valuable discussions. This work was supported by the DFG research unit 960 "Quantum phase transitions".

\newpage


\section*{Supplementary material}

**Supplemental material to „Tuning low-energy scales in YbRh$_2$Si$_2$" by M.-H. Schubert, Y. Tokiwa, S.-H. Hübner, M. Mchalwat, E. Blumenröther, H.S. Jeevan, and P. Gegenwart**

Below, we provide information on single crystal growth and characterization, electrical resistivity, heat capacity and magnetic susceptibility, magnetoresistance, electrical resistivity measurements under hydrostatic pressure, the T* crossover for fields parallel to the c-axis and the temperature dependence of the FWHM.

**Single crystal growth and characterization**

All studied single crystals were grown from In flux at temperatures between 1500 and 1000°C as described earlier [6] and characterized by powder XRD and microprobe EDX [S1]. For the latter, the crystals were polished on naturally grown ab planes, resulting in stripy surfaces as shown in Fig. S1. High precision EDX analysis confirmed homogeneous compositions. EDX on a nominally 5% Fe-doped single crystal, revealed that the Fe content can be determined with 1at-% accuracy. For Ni substituted crystals, also wavelength dispersive X-ray spectroscopy and laser ablation inductively coupled plasma mass spectrometry was utilized [S2]. Within the resolution of these techniques, all crystals are single phase. Actual doping concentrations have been determined with precision of Δx=0.01.

|  | *Fe-K* | *Rh-L* |
|---|---|---|
| *MM110504Aa(7)_pt1* | 5.81 | 94.19 |
| *MM110504Aa(7)_pt2* | 4.76 | 95.24 |
| *MM110504Aa(7)_pt3* | 5.07 | 94.93 |
| *MM110504Aa(7)_pt4* | 4.59 | 95.41 |

Atom % Error

|  | *Fe-K* | *Rh-L* |
|---|---|---|
| *MM110504Aa(7)_pt1* | +/-0.96 | +/-0.89 |
| *MM110504Aa(7)_pt2* | +/-0.52 | +/-0.91 |
| *MM110504Aa(7)_pt3* | +/-1.00 | +/-0.92 |
| *MM110504Aa(7)_pt4* | +/-0.53 | +/-0.92 |

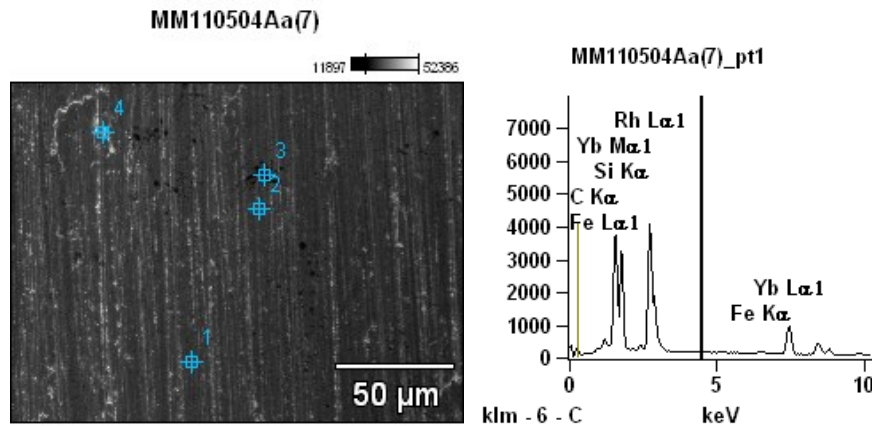

Fig. S1: Characterization of a Yb(Rh$_{0.95}$Fe$_{0.05}$)$_2$Si$_2$ single crystal by energy dispersive x-ray (EDX) analysis. Four different spots on the polished surface (left) were studied. The right figure displays data on spot #1. The tables summarize the obtained Fe- and Rh contents on the different spots and their error bars [S1].

The following two tables provide an overview on the physical properties of all Fe- and Ni-substituted crystals, studied in this work [S1-S3]. Here, $T_{max}$, has been determined from electrical resistivity, $T_K$ from the magnetic entropy, $\Theta_W$ from magnetic susceptibility measurements between 2 and 4 K and $T_N$ from electrical resistivity and specific heat measurements. Respective data for Co- and Ir-substitutions can be found in [S4-S6,9]. "-": not measured.

| x (Fe) ±0.01 | a (Å) ±0.002 | c (Å) ±0.004 | $V_{uc}$ (Å³) ±0.25 | $\rho_0$ (μΩcm) | $T_{max}$ (K) ±10 K | $T_K$ (K) ±5K | $\Theta_W$ (K) ±0.5 K | $T_N$ (mK) ±10 |
|---|---|---|---|---|---|---|---|---|
| 0 | 4.011 | 9.861 | 158.6 | 0.3-3 | 140 | 25 | −3 | 70 |
| 0.03 | - | - | - | 1.8 | 131 | 18 | −4.2 | 50 |
| 0.05 | 4.007 | 9.846 | 158.1 | - | 113 | 17 | - | 0 |
| 0.075 | - | - | - | 7.6 | 107 | - | −6.25 | 0 |
| 0.1 | 4.000 | 9.823 | 157.2 | 12.5 | 81 | 12 | −5.4 | 0 |

| x (Ni) ±0.01 | a (Å) ±0.002 | c (Å) ±0.004 | $V_{uc}$ (Å³) ±0.25 | $\rho_0$ (μΩcm) | $RR_{2K}$ | $T_{max}$ (K) ±10 K | $T_K$ (K) ±5K | $\Theta_W$ (K) ±0.5 K | $T_N$ (K) |
|---|---|---|---|---|---|---|---|---|---|
| 0.02 | - | - | - | 8.5 | 5 | 110 | 22 | −4.6 | 0.115±0.01 |
| 0.025 | 4.0076 | 9.857 | 158.32 | - | 3.3 | 100 | 17 | - | 0.17±0.02 |
| 0.044 | 4.0060 | 9.849 | 158.07 | - | 2.6 | 92 | 14 | −3.3 | - |
| 0.06 | - | - | - | 27.5 | 1.9 | 80 | - | −1.5 | 1.4±0.2 |
| 0.12 | - | - | - | - | 1.4 | 60 | 8 | −1.2 | 1.9±0.3 |
| 0.16 | 3.9896 | 9.826 | 9.826 | - | 1.3 | 55 | 7 | −0.8 | 2.2±0.3 |

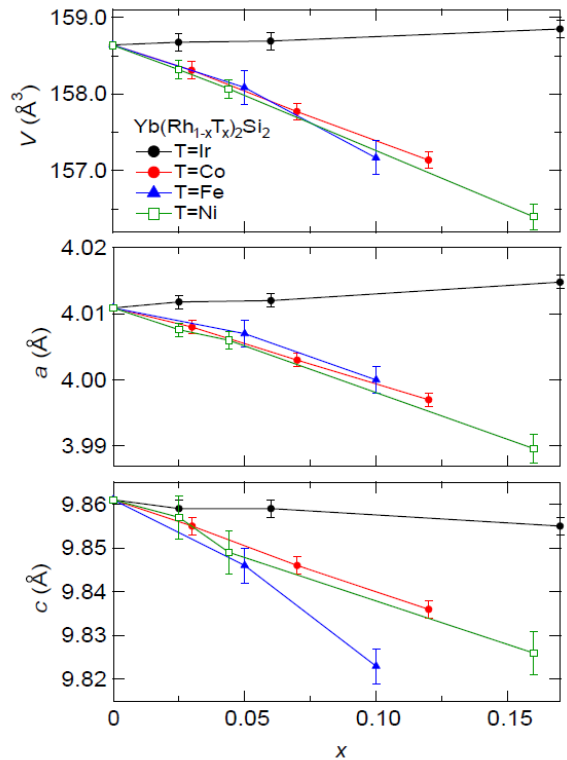

Fig. S2 Evolution of the unit-cell volume, a- and c-parameters from powder XRD on selected Yb(Rh$_{1-x}$T$_x$)$_2$Si$_2$ crystals with T=Ir [S5], Co [S6,S7], Fe [S1] and Ni [S2].

Fig. S2 displays the unit-cell volume, a- and c-axis parameters for the studied Fe- and Ni-substituted systems in comparison with literature data [32,S5] for Co- and Ir-substitution. Clearly, the evolution of the unit-cell volume for partial Fe, Co and Ni substitution of Rh decreases, while it increases for Ir-substitution, in line with the expected positive and negative chemical pressure effect, respectively.

Interestingly, the c-parameter for Yb(Rh$_{0.9}$Fe$_{0.1}$)$_2$Si$_2$ is already smaller than in pure YbFe$_2$Si$_2$ [S7-S9]. This implies a non-monotonic c(x)-dependence for x>0.1. There are further indications for drastic changes beyond x=0.1: magnetic order must reappear, since pure YbFe$_2$Si$_2$ orders below 0.75 K [30] and the ground-state crystal electric field wave function must change from $\Gamma_7$ in YbRh$_2$Si$_2$ [S10] to $\Gamma_6$ in YbFe$_2$Si$_2$ [30]. The smooth evolution of the low-temperature magnetic susceptibility and the lattice parameters up to x=0.1 indicate, that these changes appear beyond x=0.1.

**Electrical resistivity**

Fig. S3 displays $\rho$(T) of various Fe-and Ni-substituted crystals. The temperature, at which the electrical resistivity passes a maximum, $T_{max}$(x), decreases monotonically for all substitutions, which can be associated to a chemical pressure effect, since previous hydrostatic pressure experiments [S11] found a similar behavior, while for Ir-substitution (negative chemical pressure), $T_{max}$ increases [S4].

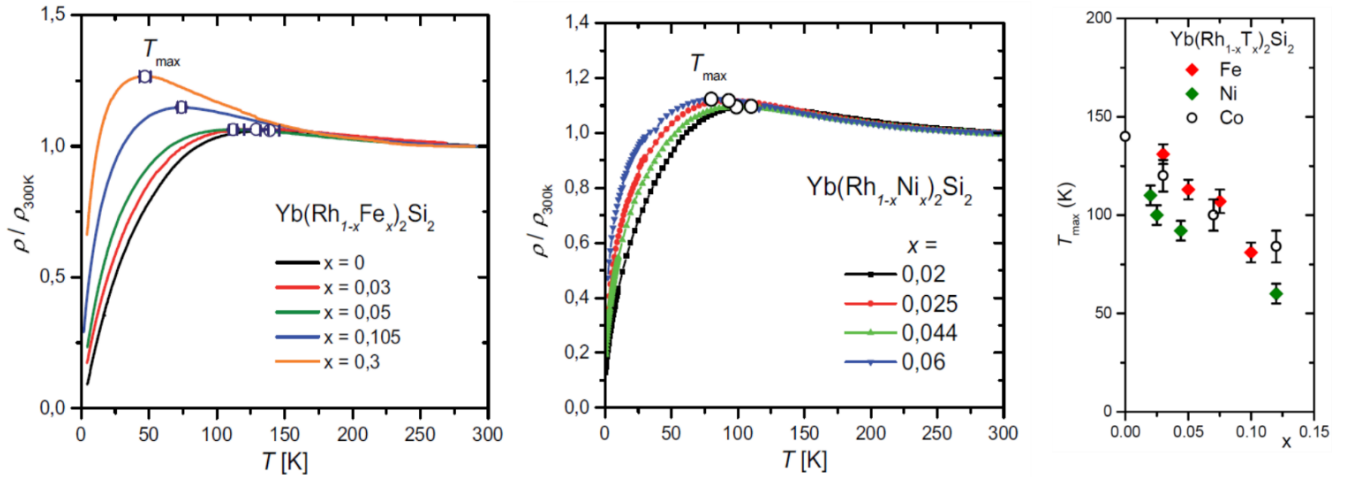

Fig. S3: Zero-field electrical resistivity of Yb(Rh$_{1-x}$T$_x$)$_2$Si$_2$ for T=Fe (left) and T=Ni (middle) [S1-S3]. The right panel displays the maximum temperature for T=Fe and Ni, in comparison with T=Co [S4].

**Specific heat**

Low-temperature specific heat measurements of various Fe- and Ni-substituted systems are shown in Fig. 1 of the main manuscript. The Kondo temperature $T_K=2\cdot T(S=0.5\ R\log 2)$, derived from magnetic entropy is listed in the tables above and shown in Fig. S4. Here we use a value of 25 K for undoped YbRh$_2$Si$_2$ [S11]. The decrease of $T_K$(x) is in accordance with effect of chemical pressure and resembles the behavior found for Yb(Rh$_{1-x}$Co$_x$)$_2$Si$_2$ [S6]. This also indicates, that the disparate dependence of $T_N$(x), which decreases for Fe-substitution but increases for Co- and Ni-substitution is related to the effect of non-isoelectronic substitution, i.e., hole-doping in Yb(Rh$_{1-x}$Fe$_x$)$_2$Si$_2$.

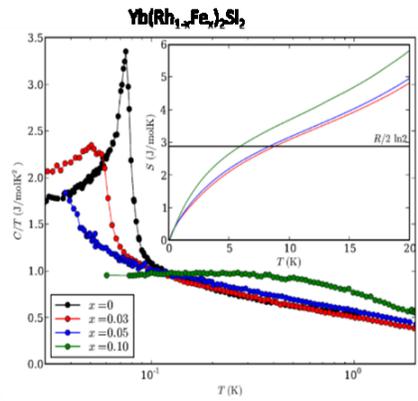 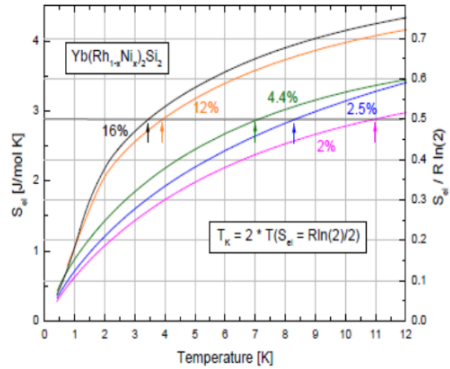 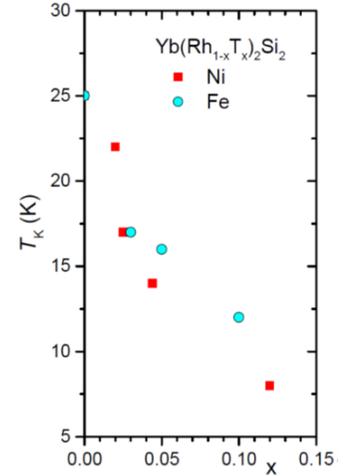

Fig. S4: Left: Heat capacity of Yb(Rh$_{1-x}$Fe$_x$)$_2$Si$_2$ and magnetic entropy (inset). Middle: magnetic entropy of Yb(Rh$_{1-x}$Ni$_x$)$_2$Si$_2$. Right: derived Kondo temperatures, determined from $T_K = 2T(S=0.5\, R\log 2)$ [S1-S3,S10].

**Magnetic susceptibility**

A similar difference between Fe- and Co or Ni substitution is also found in the Curie-Weiss temperature, obtained by fitting the low-temperature (2-4 K) in-plane magnetic susceptibility. As shown in Fig. S5, the effect of hole-doping in Fe and Ru dominates over the chemical pressure effect and tunes the Weiss temperature towards more negative values, indicating a shift of the nature of the magnetic fluctuations from ferro- to antiferromagnetic, compatible with the strong depression of $\chi(T)$ below 1 K, shown in Fig. 2 of the main text.

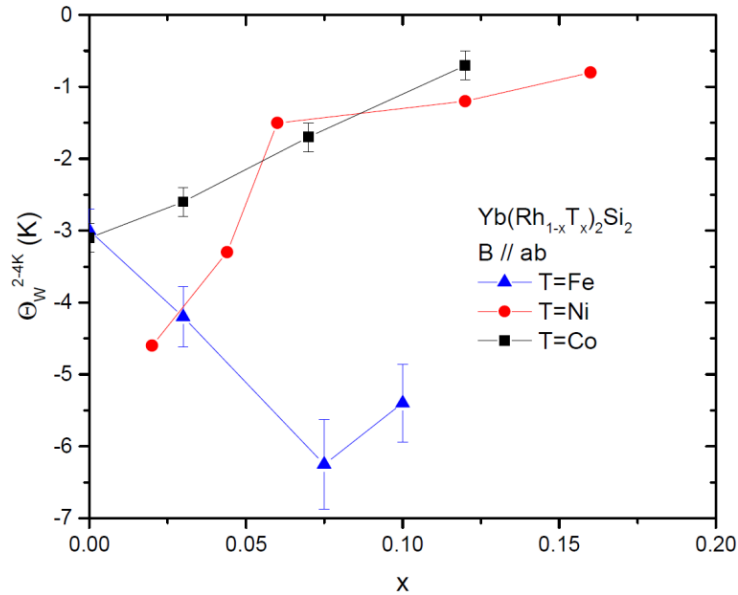

Fig. S5: Evolution of the Curie-Weiss temperature as determined from $\chi \sim (T-\Theta_W)^{-1}$ fits of the magnetic susceptibility of Yb(Rh$_{1-x}$T$_x$)$_2$Si$_2$ single crystals for T=Co [S6], Ni [4] and Fe between 2 and 4 K [S1-S3].

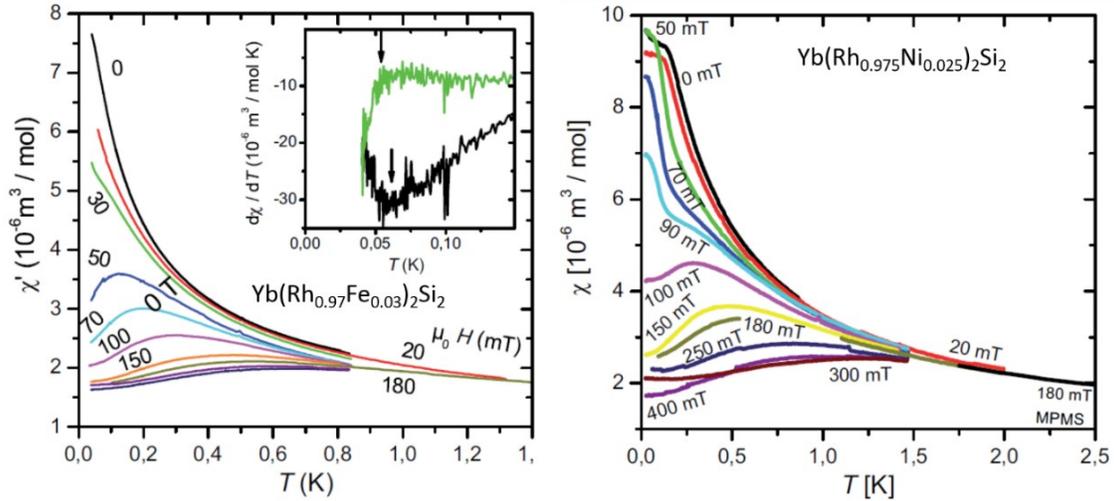

Fig. S6: Temperature dependence of the real part of the ac-susceptibility, representing the differential susceptibility dM/dB of Yb(Rh$_{0.97}$Fe$_{0.03}$)$_2$Si$_2$ (left) and Yb(Rh$_{0.975}$Ni$_{0.025}$)$_2$Si$_2$ (right) for various superposed magnetic fields indicated by the labels [S3].

AC-susceptibility measurements were performed in a homemade coil setup [S4]. Thermal coupling was achieved by glued silver wires. The ac-susceptibility was detected at a frequency near 100 Hz. Its real part represents the differential susceptibility dM/dB. To obtain SI units, the ac-susceptibility data were scaled at a temperature of 1.8 K to respective M/B data determined in a SQUID magnetometer, similar as done previously [6]. Fig. S6 displays data on two differently doped single crystals. Similar as found in previously in undoped YbRh$_2$Si$_2$ [10], the maxima in $\chi(T)$ at different fields follow T*(B).

**Magnetoresistance**

Fig. S7 displays magnetoresistance data on various studied single crystals. We analyzed all magnetoresistance data by the empirical crossover function from Ref. [14]. The Yb(Rh$_{1-x}$Fe$_x$)$_2$Si$_2$ system with x=0.085 is the last one, for which the crossover could be detected, since it is absent for x=0.1.

Fig. S8 displays for x=0.085 at the lowest measured temperatures, the magnetoresistance data (symbols) and the fit by the empirical crossover function from Paschen et al. (Ref. 14 main text). Note, the good description of the data by the fit, which still allows an accurate determination of the crossover field B* as well as crossover full width at half maximum (FWHM). The comparison of data at 15 and 30 mK evidences, that there is no proportionality of the FWHM with temperature, but rather a saturation of the latter as T→0.

The left part of Fig. S9 compares B*(T) for different Fe-substitutions. A systematic shift towards smaller fields is found. However, even for x=0.085, for which the size of the crossover is almost suppressed, B*(T=0) still amounts to 0.025 T. The x=0.1 system is paramagnetic and displays a Fermi liquid ground state (Fig. S9 right).

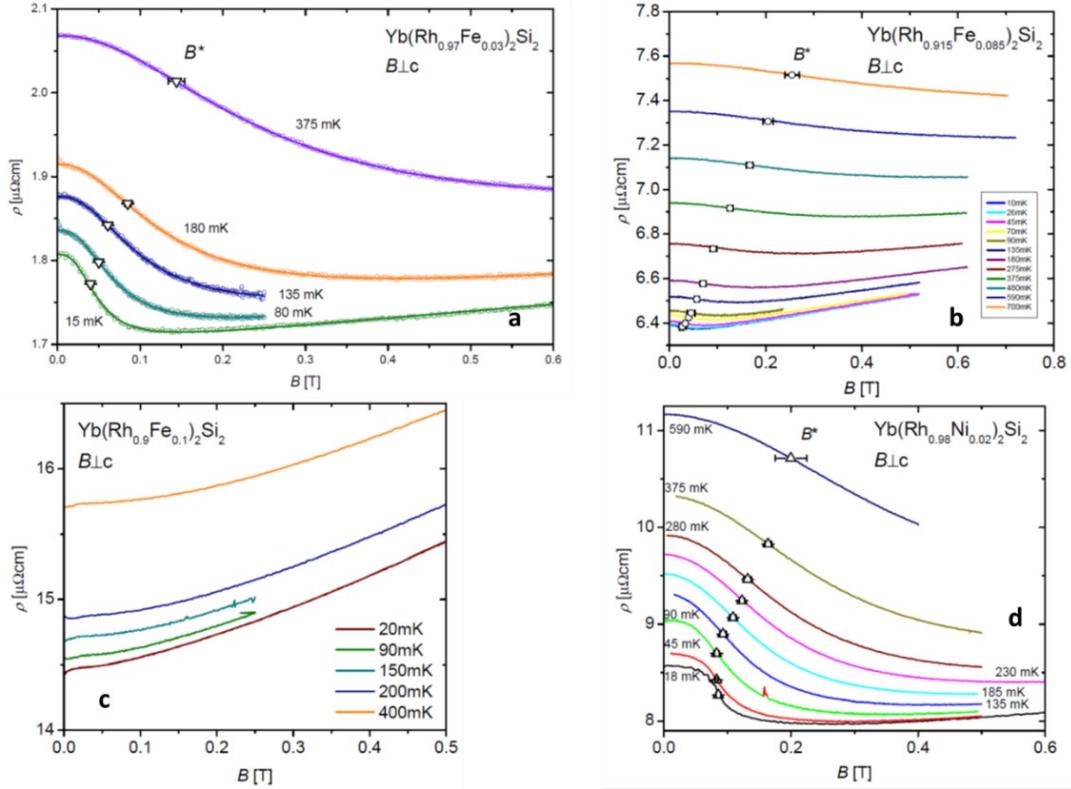

Fig. S7: Isothermal magnetoresistance measurements on various Yb(Rh$_{1-x}$T$_x$)$_2$Si$_2$ single crystals. The position of the symbols indicates the crossover field B*(T) [S3].

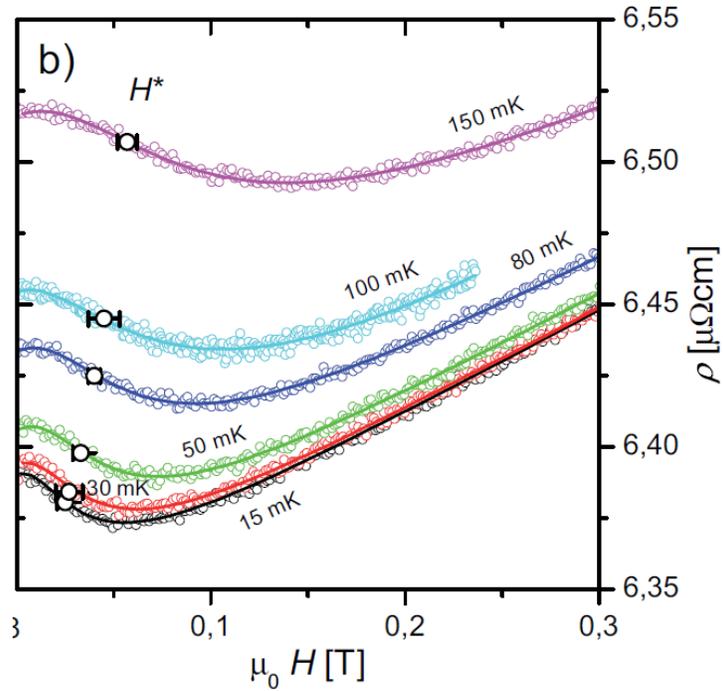

Fig. S8: Detailed view on the low-temperature isothermal magnetoresistance data of Yb(Rh$_{1-x}$T$_x$)$_2$Si$_2$, x=0.085 (cf. Fig. S7b). The symbols display the magnetoresistance data, while the lines indicate fits with the empirical crossover function (Ref. 14, main text), used to determine B* as well as the FWHM. From [S3].

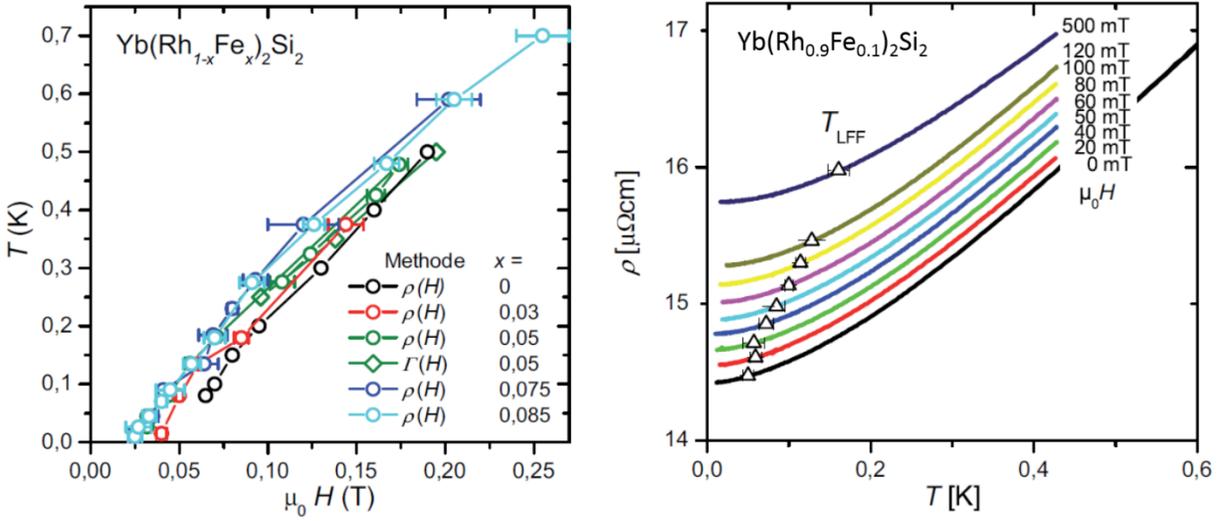

Fig. S9: Temperature vs field phase diagram with T* line for various Yb(Rh$_{1-x}$Fe$_x$)$_2$Si$_2$ single crystals with x≤0.085 (left) and temperature dependence of the electrical resistivity for Yb(Rh$_{0.9}$Fe$_{0.1}$)$_2$Si$_2$ at various magnetic fields (right); curves for different fields are shifted with respect to the zero-field data. The triangles indicate the boundary of T$^2$ Landau Fermi liquid (LFL) behavior [S3].

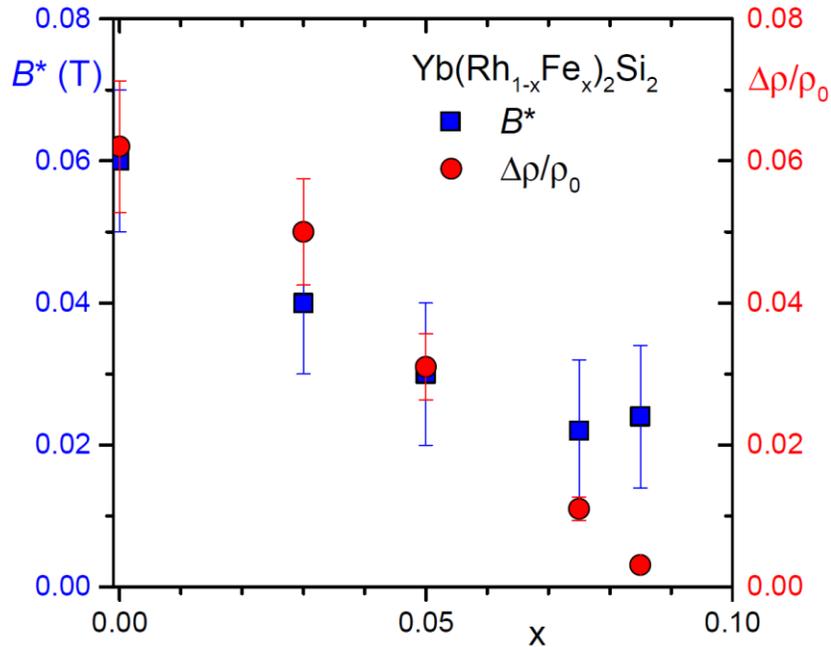

Fig. S10: Crossover field B* (blue squares) and size of magnetoresistance crossover as Δρ/ρ$_0$, determined by fitting with the crossover function from [14], vs x for various Yb(Rh$_{1-x}$Fe$_x$)$_2$Si$_2$ systems. Note, that B* does not approach zero field before the crossover vanishes.

Fig. S10, displays the evolution of the crossover field B* as well as the relative size Δρ/ρ$_0$ of the magnetoresistance crossover (determined from the fit with the empirical crossover function [14]) as function of x for Yb(Rh$_{1-x}$Fe$_x$)$_2$Si$_2$. Importantly, the magnetoresistance crossover disappears (i.e. the size extrapolates to zero) while the crossover field B* remains finite. This proves that the crossover is a field-

induced effect and does not appear without applied field. Note, that by contrast Ni-substitution enhances the relative size of the crossover (cf. Fig. S9).

**Electrical resistivity measurements under hydrostatic pressure**

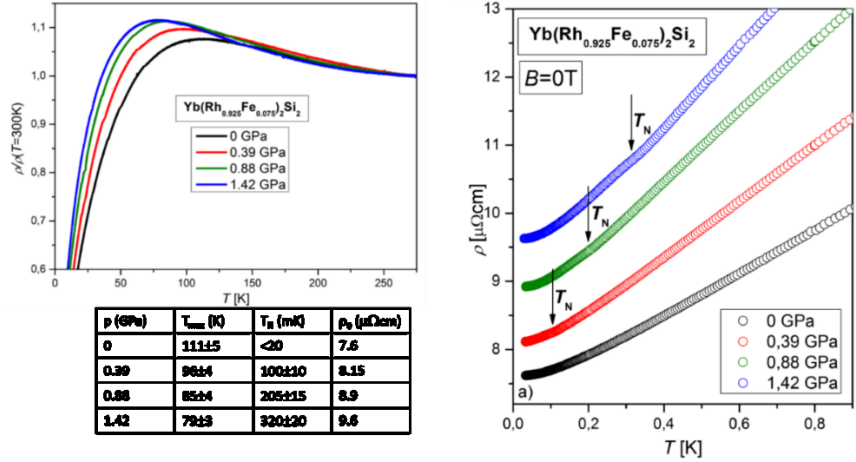

Fig. S11: Temperature dependence of the electrical resistivity of Yb(Rh$_{0.925}$Fe$_{0.075}$)$_2$Si$_2$ under various hydrostatic pressures. The table specifies the temperatures of the maximum and Néel ordering as well as the residual resistivity [S13].

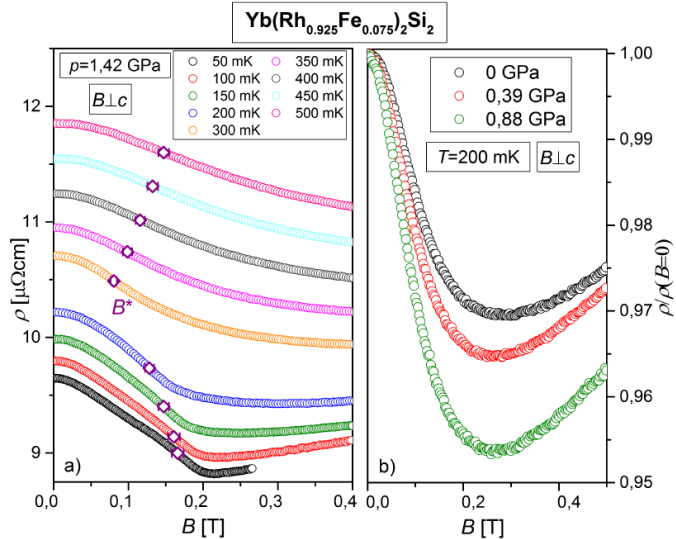

Fig. S12: Isothermal magnetoresistance of Yb(Rh$_{0.925}$Fe$_{0.075}$)$_2$Si$_2$ under hydrostatic pressure of 1.42 GPa (left) and various different pressures (right). [S13].

We first discuss Yb(Rh$_{0.925}$Fe$_{0.075}$)$_2$Si$_2$, cf. Fig. S11. The resistivity maximum shifts with increasing pressure to lower values. Below 1 K, $\rho(T)$ displays characteristic anomalies associated with the onset of magnetic order. As shown in the inset of Fig. 5 in the main part of the manuscript, $T_N$ has been determined from a dip in the temperature derivative of the electrical resistivity [S13], similar as done previously for Co-substituted YbRh$_2$Si$_2$, cf. Fig. S13 [S14]. Note, that these values agree with $T_N$ from ac-susceptibility [S3]. On the other hand, $B^*$ for Co-doped YbRh$_2$Si$_2$ has been determined from isothermal magnetoresistance for $T>T_N$. For $T<T_N$ (cf. Fig. S13 left), the inflection point of the magnetoresistance follows the AF phase boundary and $B^*$ cannot any more be determined from magnetoresistance [S14]. Here the $T^*(B)$ crossover was determined inside the AF state from ac-susceptibility measurements [9].

A full set of magnetoresistance data for the highest applied pressure on Yb(Rh$_{0.925}$Fe$_{0.075}$)$_2$Si$_2$ is shown in Fig. S12 left. The symbols indicate the inflection points. The right panel shows a comparison of the normalized magnetoresistance crossover for various pressures at a temperature of 200 mK.

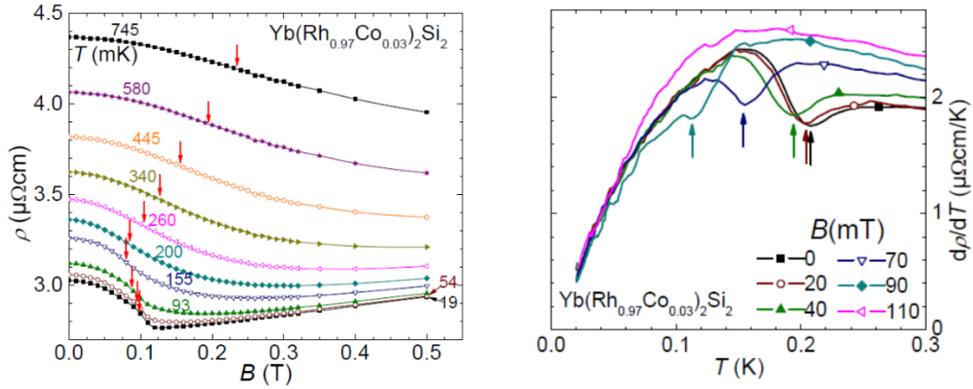

Fig. S13: Isothermal magnetoresistance of Co-doped YbRh$_2$Si$_2$ [S14] (left). The red arrows mark inflection points. The fall red arrows mark inflection points. Temperature derivative of the electrical resistivity (left). Arrows mark mimima, used to determine $T_N$ [S14].

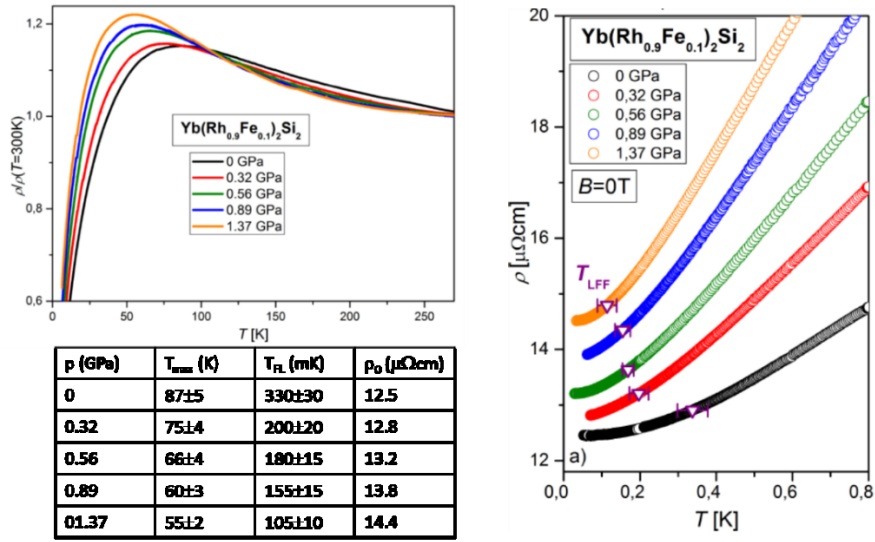

| p (GPa) | T$_{max}$ (K) | T$_{FL}$ (mK) | ρ$_0$ (μΩcm) |
|---|---|---|---|
| 0 | 87±5 | 330±30 | 12.5 |
| 0.32 | 75±4 | 200±20 | 12.8 |
| 0.56 | 66±4 | 180±15 | 13.2 |
| 0.89 | 60±3 | 155±15 | 13.8 |
| 01.37 | 55±2 | 105±10 | 14.4 |

Fig. S14: Temperature dependence of the electrical resistivity of Yb(Rh$_{0.9}$Fe$_{0.1}$)$_2$Si$_2$ under various hydrostatic pressures. The table specifies the temperatures of the maximum and T$^2$ boundary as well as the residual resistivity. [S13].

We now turn to the hydrostatic pressure effect on Yb(Rh$_{0.9}$Fe$_{0.1}$)$_2$Si$_2$ [S13]. Fig. S14 displays zero-field electrical resistivity data at various pressures. The left panel indices the shift of the resistivity maximum with pressure, while the right panel shows the data below 1 K at various pressures. The triangles mark the temperatures $T_{FL}$, below which Fermi liquid behavior $\Delta\rho \sim T^2$ is observed. With increasing pressure, $T_{FL}$ shifts towards zero but the maximal pressure is too small to suppress the Fermi liquid ground state.

Finally, we show the pressure dependence of the isothermal magnetoresistance of Yb(Rh$_{0.9}$Fe$_{0.1}$)$_2$Si$_2$ in the left panel of Fig. S15. An increase of the electrical resistance with field is found, which is even enlarged with hydrostatic pressure. Even at elevated temperatures (right panel) no negative magnetoresistance is found. Thus, the T* anomaly is completely absent in Yb(Rh$_{0.9}$Fe$_{0.1}$)$_2$Si$_2$.

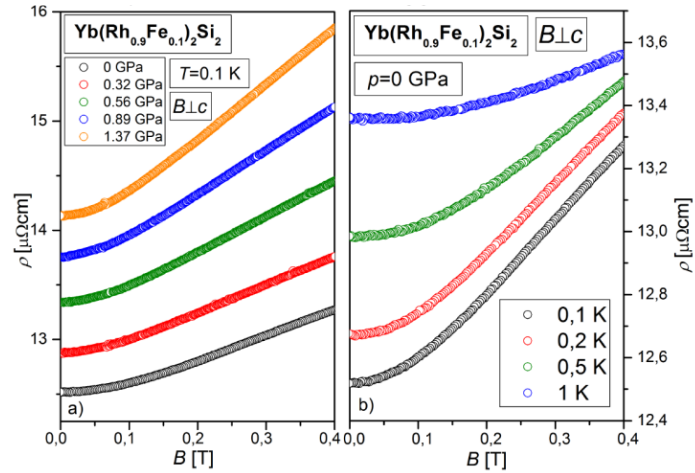

Fig. S15: Isothermal magnetoresistance of Yb(Rh$_{0.9}$Fe$_{0.1}$)$_2$Si$_2$ under various hydrostatic pressures at 0.1 K (left) and for ambient pressure at various temperatures (right). Data in the right panel have been shifted by constant resistivity values with respect to each other [S13].

**T* crossover for fields parallel to the c-axis**

In our study, the magnetic field has always been applied perpendicular to the c-axis, i.e., within the easy magnetic plane (ab). As shown in [16], the T* crossover from magnetoresistance perpendicular to the c-axis perfectly agrees with that from crossed-field Hall measurements. In the latter case, the Hall effect was induced by a small field along the c-axis (hard axis) and an in-plane field parallel to the applied current was used to tune the ground state from AF to paramagnetic at low temperatures. T* for fields along the c-axis (hard axis) has only been studied in single-field Hall measurements. Here, the slope of the Hall voltage was analyzed. The resulting T*(B) perfectly agrees with that determined for B perpendicular c, of the magnetic anisotropy factor 11 is used to scale the field axis. This suggests, that the T*(B) crossover determined for both field orientations has similar origin, although the role of ferromagnetic correlations for B//c is unclear, because yet there exist no reliable low-temperature magnetization measurements along the hard axis.

**Temperature dependence of the FWHM**

In [15] the isothermal magnetoresistance for fields perpendicular to the c-axis was only analyzed above 100 mK, since below an additional maximum appeared, related to the AF phase boundary, which disturbed the analysis. Ref. [16] reported a more detailed study of the temperature dependence of the FWHM (Fig. S16). Here, magnetoresistance of sample 1 does not show this feature at the AF phase boundary. Nevertheless, the derived FWHM (crosses in Fig. S16e) clearly deviates from the linear dependence (red line), expected within the Kondo breakdown scenario.

In fact, all previous FWHM data points below 40 mK deviate from the red line (see Fig. S16b), with the exception of the *two open circles* (note in the magnified plot b, that the lower error bar of the lowest circle infers with a small axis tick, looking like a plus symbol below the red line, however, this is not a true data point). These two open circles were determined from single-field Hall measurements on sample 2.

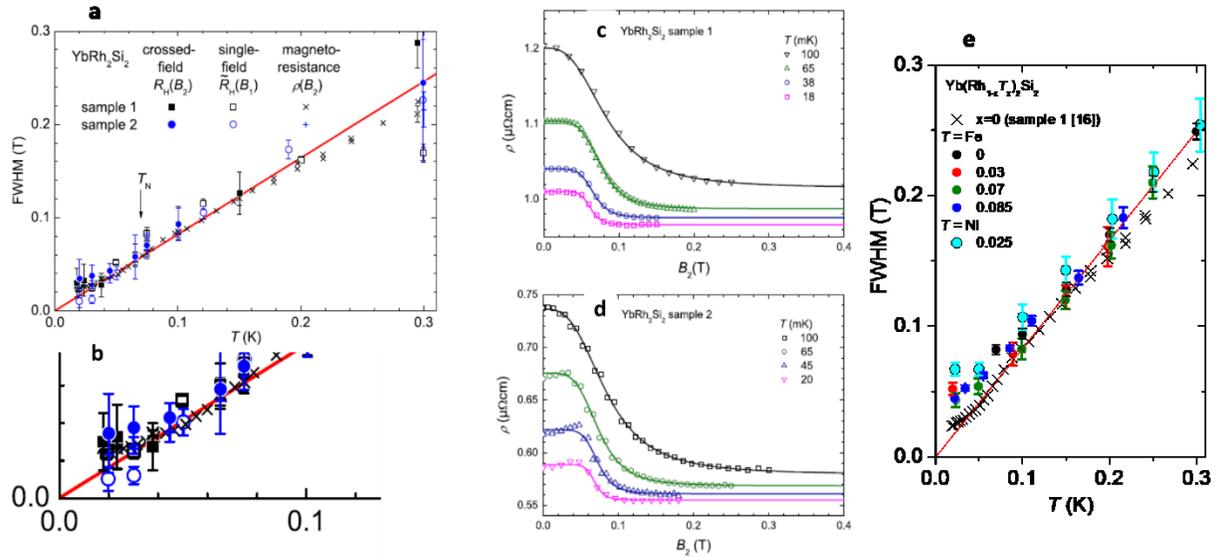

Fig. S16: Magnetoresistance (c,d) and FWHM (a,b) from Ref. [16]. (e): Comparison of FWHM from magnetoresistance of various doped Yb(Rh$_{1-x}$T$_x$)$_2$Si$_2$ systems with x=0 from sample 1 from [16]; red line similar as in [16].

However, as shown in Fig. S4 of [16], the differential Hall coefficient dR$_H$/dB(B$_1$), derived from single-field Hall measurements, does not properly follow the expected crossover function, questioning the reliability of the two data points *below the red line*.

This indicates already in [16] a discrepancy to the Kondo breakdown interpretation. In our study, magnetoresistance crossovers have only been analyzed at temperatures above the respective T$_N$ of various doped systems, in order to avoid any interference of the T* crossover with the AF phase boundary. Clearly, the FWHM remains finite in the zero-temperature limit (Fig. S16e).

**Supplemental References**


\end{document}